%% file: main.tex
\documentclass[aps,prl,reprint,
							floatfix,
							superscriptaddress,
							footinbib,
							floatfix,
							longbibliography]{revtex4-1}

\usepackage{amsmath}
\usepackage{amssymb}
\usepackage[normalem]{ulem}

\usepackage{natbib}
\usepackage[T1]{fontenc}
\usepackage[latin1]{inputenc}

\usepackage[dvipsnames]{xcolor}

\usepackage{hyperref}
\hypersetup{colorlinks=true,citecolor=Cyan,urlcolor=Red}

\usepackage[squaren]{SIunits}

\usepackage[pdftex]{graphicx}
\usepackage{grffile} 
\usepackage{subfigure}

\let\vec\boldsymbol

\newcommand{\RefA}[1]{{#1}}
\newcommand{\RefB}[1]{{#1}}

\newcommand{\ud}{\mathrm{d}}

\begin{document}

\title{Optimizing Laser Pulses for Narrowband Inverse Compton Sources in the High-Intensity Regime}

\author{Daniel~Seipt}
\email{dseipt@umich.edu}
\affiliation{Helmholtz-Institut Jena, Fr\"obelstieg 3, 07743 Jena, Germany}
\affiliation{Center for Ultrafast Optical Science, University of Michigan, Ann Arbor, Michigan 48109, USA}
 
\author{Vasily~Yu.~Kharin}
\affiliation{Helmholtz-Institut Jena, Fr\"obelstieg 3, 07743 Jena, Germany}

\author{Sergey~G.~Rykovanov}
\email{s.rykovanov@skoltech.ru}
\affiliation{Helmholtz-Institut Jena, Fr\"obelstieg 3, 07743 Jena, Germany}
\affiliation{Center for Computational and Data-Intensive Science and Engineering, 
Skolkovo Institute of Science and Technology, Nobel Str. 3, Skolkovo, Russia}

\begin{abstract}
Scattering of ultraintense short laser pulses off relativistic electrons allows one to generate a large number of X- or gamma-ray photons with the expense of the spectral width---temporal pulsing of the laser inevitable leads to considerable spectral broadening. In this Letter, we describe a simple method to generate optimized laser pulses that compensate the nonlinear spectrum broadening, and can be thought of as a superposition of two oppositely linearly chirped pulses delayed with respect to each other. We develop a simple analytical model that allow us to predict the optimal parameters
of such a two-pulse---the delay, amount of chirp and relative phase---for generation of a narrowband $\gamma$-ray spectrum. Our predictions are confirmed by numerical \RefA{optimization and simulations including 3D effects.}
\end{abstract}

\date{\today}

\maketitle

    The Inverse Compton Scattering (ICS) of laser light off high-energy electron beams is a well-established source of X- and gamma-rays for applications in medical, biological, nuclear, and material sciences \cite{HIGS,Bertozzi:PRC2008,Albert:PRSTAB2010,Albert:PRSTAB2011,Quiter:NIMB2011,Rykovanov:JPB2014,Albert:PPCF2016,ELI-NP,Kramer:SciRep2018}. \RefB{One of the main advantages of ICS photon sources is the possibility to generate narrowband MeV photon beams, as opposed to a broad continuum of Bremsstrahlung sources, for instance. The radiation from 3rd and 4th generation light-sources, on the other hand, typically has their highest brightness at much lower photon energies \cite{Albert:PRSTAB2010}, giving ICS sources their unique scope \cite{Nedorezov:PhysUsp2004,Carpinelli:NIMA2008,Albert:PPCF2014}.}

    With laser plasma accelerators (LPA), stable GeV-level electron beams have been produced with very high peak currents, small source size, intrinsic short duration (few 10s fs) and temporal correlation with the laser driver \cite{Mangles:Nature2004,Leemans:Nature2006,Esarey:RevModPhys2009}, which is beneficial for using them in \RefB{compact all-optical} ICS. In particular the latter properties allow for the generation of femtosecond gamma-rays that can be useful in time-resolved (pump-probe) studies.
    For designing narrowband sources, it is important to understand in detail the different contributions to the scattered radiation bandwidth \cite{Rykovanov:JPB2014,Curatolo:PRSTAB2017,Kramer:SciRep2018,Ranjan:PRSTAB2018}.

    For an intense source the number of electron-photon scattering events needs to be maximized \cite{Rykovanov:JPB2014}, which can be achieved in the most straightforward way by increasing the intensity of the scattering laser, $I[\watt\per\centi\metre^2] = 1.37\times 10^{18}\, a_0^2 / \lambda^2[\micro\metre]$, where $a_0=e A_0/m$ with $e$ and $m$ electron absolute charge and mass respectively, and $A_0$ the \RefB{peak} amplitude of the laser pulse vector potential in Gaussian CGS units (with $\hbar=c=1$) and $\lambda$ the laser wavelength. However, this leads to an unfortunate consequence when the laser pulse normalized amplitude reaches $a_0\sim 1$: ponderomotive spectral broadening of the scattered radiation on the order of $\Delta\omega' /\omega' \sim a_0^2/\left(1+a_0^2 \right)$  \cite{Hartemann:PRE1996,Krafft:PRL2004,Hartemann:PRL2010,Maroli:PRSTAB2013, Rykovanov:JPB2014,Curatolo:PRSTAB2017}.

    The ponderomotive broadening is caused by the \RefA{$\vec v\times \vec B$
    force, which effectively slows down the forward motion of electrons near the peak of the laser pulse where the intensity is high} \cite{Krafft:PRL2004,Hartemann:PRL2010,Kharin:PRA2016}. Hence, the photons scattered near the peak of the laser pulse are red-shifted compared to the photons scattered near the wings of the pulse and the spectrum of gamma rays becomes broad. Broadband gamma-ray spectra from ICS using laser-accelerated electrons and scattering pulses with $a_0>1$ have been observed experimentally \cite{Chen:PRL2013,Sarri:PRL2014,Khrennikov:PRL2015,Cole:PRX2018}.

    To overcome this fundamental limit it was proposed to use temporal laser pulse chirping to compensate the nonlinear spectrum broadening  \cite{Ghebregziabher:PRSTAB2013,Terzic:PRL2014,Seipt:PRA2015,Rykovanov:PRSTAB2016,Kharin:PRL2018}. By performing a stationary phase analysis of the non-linear Compton S-matrix elements \cite{Seipt:PRA2015} or the corresponding classical Li\'{e}nard-Wiechert amplitudes \cite{Terzic:PRL2014,Rykovanov:PRSTAB2016} one finds that the fundamental frequency of on-axis ICS photons behaves as (off-axis emission and higher harmonics can be easily accounted for \cite{Seipt:PRA2015}).
    \begin{align}
    \omega'(\varphi) 
        = 
        \frac{4 \gamma^2 \omega_L }{1 + a(\varphi)^2 + \frac{2\gamma\omega_L}{m}} \,,
    \end{align}
    where the last term in the denominator is the quantum recoil and missing in a classical approach. Here $a(\varphi)$ is the envelope of the normalized laser pulse vector potential, with peak value $a_0$, $\omega_L$ as the laser frequency, and $\varphi = t - z$. If the instantaneous laser frequency follows the laser pulse envelope according to
    \begin{align} \label{eq:optimum-chirp}
    \omega_L(\varphi) 
        = 
        \omega_{0} [ 1 + a(\varphi)^2 ] \,,
    \end{align}
    with a reference frequency $\omega_{0}$, then the non-linear broadening is perfectly compensated, $\omega' = const. \approx 4\gamma^2 \omega_0$ in the recoil free limit.    This is the basic idea behind the promising approach for ponderomotive broadening compensation in ICS using chirped laser pulses. However, realizing such a highly nonlinear temporal chirp experimentally seems challenging because the laser frequency needs to precisely sweep up and down again within a few femtoseconds \cite{Terzic:PRL2014,Kharin:PRL2018}.

    In this Letter, we propose a simple method to generate optimized laser pulses for the compensation of the nonlinear broadening of ICS photon sources and, thus, enhance photon yield in a narrow frequency band. We propose to synthesize an optimized laser spectrum using standard optical dispersive elements. By working in frequency space, both the temporal pulse shape and the local laser frequency are adjusted simultaneously to fulfill the compensation condition, Eq.~\eqref{eq:optimum-chirp}, where the frequency first rises until the peak of the pulse intensity and then drops again. We develop an analytic model to predict the optimal dispersion needed for generating a narrowband ICS spectrum, and we compare with numerical optimization of the peak spectral brightness of the compensated nonlinear Compton source. \RefA{We performed simulations for realistic electron beams, and taking into account laser focusing effects.}

\begin{figure}[t]
    \begin{center}
    \includegraphics[width=0.70\columnwidth]{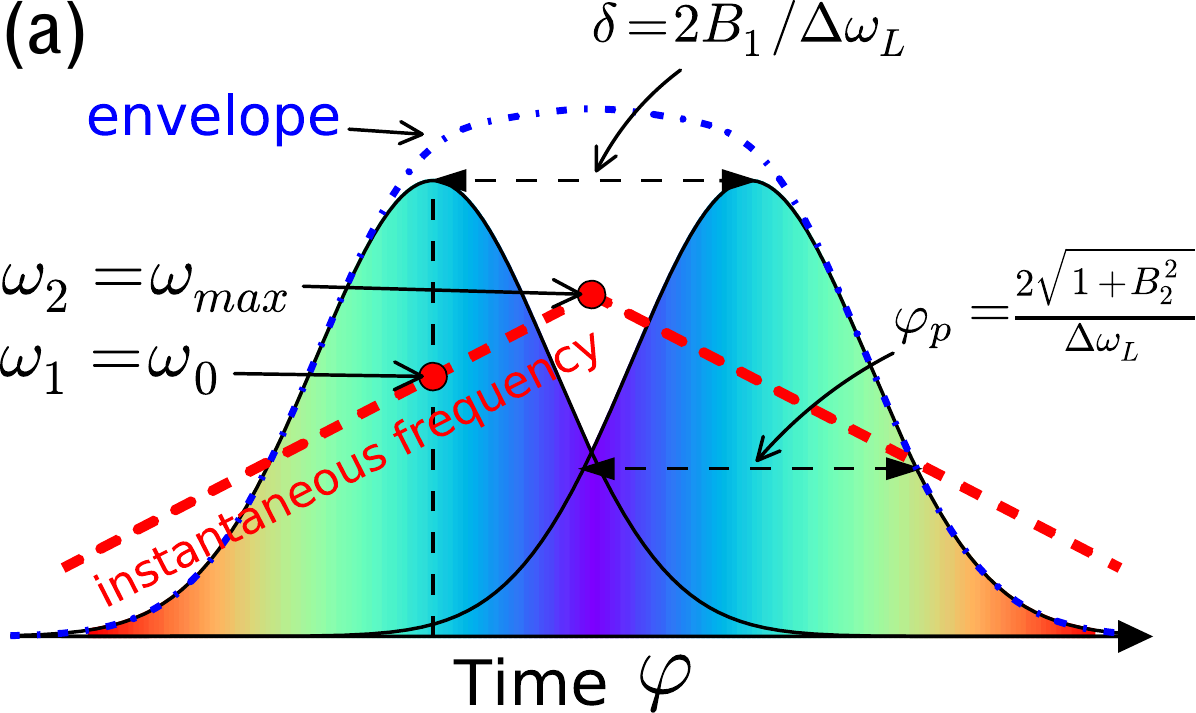}
    \includegraphics[width=0.95\columnwidth]{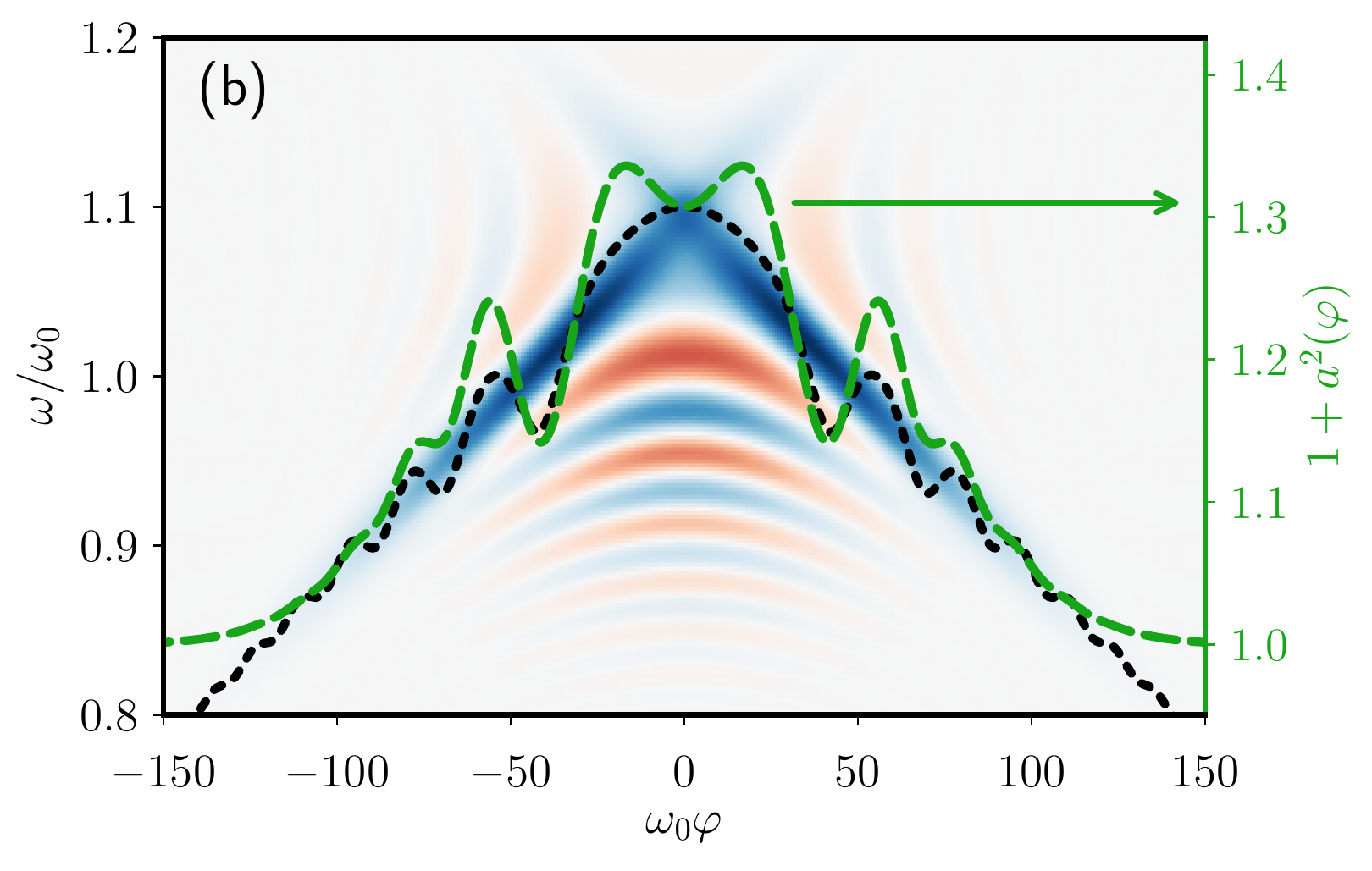}
    \end{center}
    \caption{
    \RefB{Schematic sketch of the two-pulse model for generating optimized laser pulses for narrowband nonlinear ICS (a). Wigner function of the recombined two-pulse for $\vec B = \{1.4,4.5,-4.6\}$ (b). The black curve is the analytical expression for the instantaneous frequency $\omega_L(\varphi)$, the green dashed curve is the analytical instantaneous $1+a^2(\varphi)$.}}
    \label{fig:wigner}
    \end{figure}

    In order to produce the optimized laser spectra we propose a two-pulse scheme: An initially unchirped broadband laser pulse with the spectral amplitude $\tilde a_\mathrm{in}(\omega)$ is split into two identical pulses, e.g.~using a beam splitter.  Each of these pulses is sent to the arms of an interferometer where a spectral phase $\tilde \Phi(\omega)$ is applied to one of the pulses and the conjugate spectral phase $-\tilde \Phi (\omega)$ is imposed onto the other one, using, for example, acousto-optical dispersive filters, diffraction gratings, or spatial light modulators. The two pulses are coherently recombined causing a spectral modulation. In the time-domain this translates to the coherent superposition of two linearly and oppositely chirped laser pulses which are delayed with respect to each other, see Fig.~\ref{fig:wigner}(a).

    We model the initial laser pulse by a Gaussian spectral amplitude,
    \begin{align}
    \tilde a_\mathrm{in}(\omega) 
    &=  \frac{a_0}{\Delta \omega_L}  \sqrt{2\pi} 
    \: e^{-\frac{(\omega-\omega_{0})^2}{2 \Delta\omega_L^2}} \,,
    \end{align}
    with bandwidth $\Delta\omega_L$. The inverse Fourier transform of this spectrum gives complex amplitude,
    \begin{align} \label{eq:fourier-limited}
    a_\mathrm{in}(\varphi) = a_0 
		\, e^{- \frac{\varphi^2 \Delta \omega_L^2}{2}} 
        \, e^{- i\omega_0\varphi} =  a(\varphi) \, e^{-i\omega_0\varphi} \,,
    \end{align}
    which determines the real vector potential of a circularly polarized laser pulse propagating in the $z$ direction via $\vec a_\perp = \Re [ a_\perp \vec \epsilon ]$ with $\vec \epsilon = \vec e_x + i \vec e_y$, and with $\varphi = t+z$.

    After the spectral phases $\pm \tilde \Phi(\omega)$ are applied to the two split pulses and the pulses have been recombined, the spectrum of the recombined two-pulse is modulated as
    \begin{align} \label{in_spec}
    \tilde a(\omega)  = \tilde a_\mathrm{in} (\omega) \cos \tilde \Phi(\omega) \,.
    \end{align}
    The spectral phase is parametrized as
    $\tilde \Phi(\omega) = \sum_{k=0}^2 B_k \,\left(\omega-\omega_{0}\right)^k /(k!\Delta\omega_L^k)$.
    When including higher order terms we didn't find great improvements, so we keep here only terms up to quadratic order. The dimensionless parameters $B_0$, $B_1$ and $B_2$ determine the relative phase, relative delay and amount of linear chirp, respectively, and they are collated into a vector $\vec{B}=\left\{B_0, B_1, B_2 \right\}$ for short-hand notation.

    The temporal structure of the recombined pulse can be found analytically by the inverse Fourier transform of Eq.~\eqref{in_spec}, yielding the complex amplitude $a_\perp(\varphi)=a(\varphi)e^{-i\Phi(\varphi)}$, where $a(\varphi)$ is the time-dependent envelope, and $\Phi(\varphi)$ is the temporal phase of the two-pulse. The instantaneous frequency in the recombined pulse is given by $\omega_L(\varphi) = \ud\Phi(\varphi)/\ud\varphi$, \RefB{which is a slowly varying function \cite{Maroli:JApplPhys2018}.} Explicit analytical expressions are lengthy, and given in the Supplementary Material. In the limit $\vec{B}=\vec{0}$, we re-obtain the Fourier-limited Gaussian pulse with the amplitude $a_0$ and temporal r.m.s. duration $\Delta \varphi = 1/\Delta\omega_L$, Eq.~\eqref{eq:fourier-limited}. For nonzero values of $\vec B$ the resulting laser pulse will be stretched, with a non-Gaussian envelope in general, and with a lower effective peak amplitude than the Fourier limited pulse. It can be thought of as a superposition of two interfering linearly chirped Gaussian pulses with a delay of $\delta=2B_1/\Delta\omega_L$, each with a full-width at $1/e$ duration $\varphi_p=2\sqrt{1+B_2^2}/\Delta\omega_L$,  see Fig.~\ref{fig:wigner} (a). The Wigner function characterizing such a recombined two-pulse is shown in Fig.~\ref{fig:wigner}(b), together with the instantaneous frequency $\omega_L(\varphi)$ (dashed black curve) and instantaneous $1+a^2(\varphi)$ (dashed green curve).

    The goal is to optimize $\vec B$ such that the compensation condition \eqref{eq:optimum-chirp} is fulfilled as good as possible, see green and black curves in Fig.~\ref{fig:wigner}(b). The optimal values of $\vec B$ depend on $a_0$ and the available laser bandwidth. We mention here that the stretching of the pulse alone leads to some improvement of the spectral brightness already since the effective $a_0$ is lowered. However, as we shall see later when comparing with an unchirped matched Gaussian pulse, the optimal pulse chirping improves the spectral brightness much further.

\begin{figure}
    \includegraphics[width=\columnwidth]{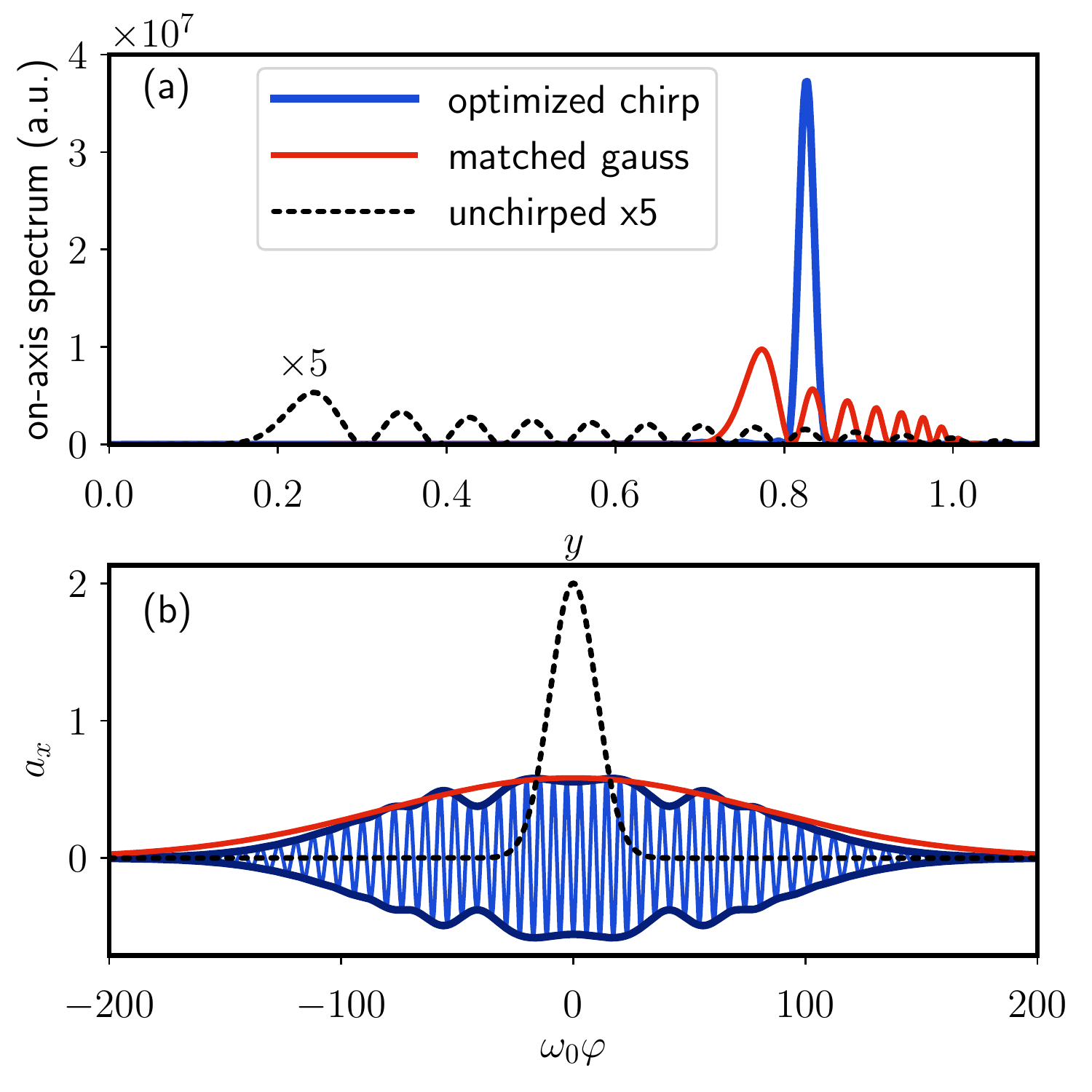}
     \caption{(a) On-axis scattered photon spectra for $a_0=2$, $\Delta\omega_L/\omega_{0}=0.1$ for different values of parameters $\vec{ B }$. The spectrum for the unchirped case with $\vec{B} = \vec{ 0 }$ is drawn with dashed black line. Note, that for visibility, the spectrum has been multiplied by 5. The case of the optimal set of parameters $\vec{ B }=\{1.4, 4.5, -4.6\}$ is shown with a blue curve. The red cure demonstrates the on-axis scattered photon spectrum for the matched unchirped Gaussian pulse (described in the text). (b) Corresponding amplitudes $a(\varphi)$ of the normalized laser vector potential.
     }\label{spectra_fig}
    \end{figure}

    The on-axis ICS gamma photon spectral density of an optimized laser pulse scattered from a counterpropagating electron with $\gamma\gg1$, assuming quantum recoil and radiation reaction can be still neglected \footnote{Quantum recoil and spin effects can be neglected for electron energies below 200 MeV, but could easily be included and do not change any qualitative findings on the optimal chirp \cite{Seipt:PRA2015}. Radiation reaction effects and recoil broadening effects due to multi-photon emission can be neglected when the number of scatterings per electron is less than one, $N_\mathrm{sc} \simeq  \frac{2\sqrt{\pi} \alpha}{3} a_0^2 \frac{\omega_0}{\Delta \omega_L} <1$, i.e. $a_0 < 3.4$ for $\Delta \omega_L/\omega_0=0.1$ \cite{Rykovanov:JPB2014,Terzic:PRL2014}. Otherwise the recoil contribution to the bandwidth is proportional to $(N_\mathrm{sc}-1) \frac{2\gamma\omega_0}{m}$.}, can then be written as \cite{Supplement}
    \RefB{
    \begin{align}
    \left.
    \frac{\ud^2 N}{\ud y \ud \Omega}
    \right|_{\theta=\pi}
	    =
	\mathcal N
    \left|\intop_{-\infty}^{+\infty} \! \ud \varphi \: 
    a_\perp(\varphi)\, 
    e^{i \omega_0 y \left(\varphi+\int_{-\infty}^{\varphi} a(\varphi')^2 \ud \varphi' \right)} \right|^2 \,,
    \label{photon_spec_eq}
    \end{align}
    with $y=\omega'/(4\gamma^2 \omega_{0})$ as the normalized frequency, $\theta$ is the polar angle,  and normalization $\mathcal N = e^2 y \gamma^2 \omega_0^2/\pi^2$.} The on-axis scattered photon spectrum depends on the values of $\vec{B}$, as illustrated in Fig.~\ref{spectra_fig}(a), where the case of $a_0=2$ and $\Delta\omega_L/\omega_{0} = 0.1 $ is presented.  The black dashed curve shows the spectrum for the unchirped laser pulse with $\vec B =\vec{ 0 }$, while the blue solid line shows the spectrum for the optimized values $\vec{B}=\{1.4, 4.5, -4.6\}$ with rms bandwidth $\Delta y = 0.0178$. 
    The corresponding normalized laser pulse vector potential shapes are shown in Fig.~\ref{spectra_fig}(b) with the same color code.

    For the chirped pulse (blue solid line), both the envelope and the waveform are presented, and one can measure that the frequency in the wings of the laser pulse is lower than in the center, where the intensity is higher. For comparison, we also employ the notion of the  matched Gaussian pulse---a pulse that has the same amplitude as the chirped pulse ($a_\mathrm{eff} = 0.58$), and the same energy content, but the frequency is constant and is equal to $\omega_{0}$ \cite{Supplement}. The envelope of the matched Gaussian pulse and the corresponding ICS photon spectrum are shown by red curves in Fig.~\ref{spectra_fig}(b) and (a), respectively. One can see that, by choosing the optimal chirp parameters, the peak of the scattered photon spectrum is in this case 4 times higher and the bandwidth is significantly narrower than the case of the matched unchirped Gaussian pulse.

    We now develop a model to predict the optimal chirp parameters $\vec B$ for generation of a narrowband ICS spectrum. As discussed above, the recombined optimized laser pulse can be seen as the coherent superposition of two delayed oppositely linearly chirped Gaussian pulses. If the delay is too large this will result in two separate pulses and the optimization condition \eqref{eq:optimum-chirp} cannot be fulfilled.

    First, for optimal pulse overlap the delay between the two pulses $\delta$ should roughly equal the duration of each of the pulses $\varphi_p$, hence, $B_1 = \chi \sqrt{1+B_2^2}$, with $\chi=\mathcal O(1)$ as a factor of proportionality \footnote{Due to symmetry reasons, $-\vec B$ produces the same laser pulse as $\vec B$. Moreover, $B_1$ and $B_2$ need to have opposite sign, we chose here $B_1>0$.}. Second, the interference of two pulses should be mostly constructive and the two-pulse should contain the maximum possible amount of photons, which determines $B_0(B_1,B_2)$, see Eq.~\eqref{B_model} below, by requiring that the argument of the cosine in the analytical expression for $a^2(\varphi)$ \cite{Supplement} is $2\pi n$ with integer $n$.

    Finally, but most importantly, we match the linear chirp given by parameter $B_2$ to the change of the envelope. For doing this, one only needs two values of instantaneous frequency $\omega_L(\phi)$ \cite{Supplement}: one at the center of each Gaussian pulses, $\omega_1=\omega_{0}$, and the other in the middle of the two-pulse, $\omega_2 \simeq \omega_{0} - {\chi B_2 \Delta\omega_L}/{\sqrt{1+B_2^2}} $. This is outlined in Fig.~\ref{fig:wigner} (a), where instantaneous frequency is schematically drawn with the dashed red line, and two points used for slope matching are shown with red dots. The compensation condition for narrowband emission now turns into
    \begin{align}
    \frac{\omega_1}{1+a^2(-\delta/2) } = \frac{\omega_2}{1+a^2(0)}\,,
    \end{align}
    which provides an equation for $B_2$. In the limit $|B_2|\gg 1$, after straightforward algebraic calculations, we find expressions for all three parameters as functions of laser bandwidth and $a_0$,
    \begin{align}
    \begin{split}
	B_2 &= -\frac{a_0^2}{4\chi}\frac{\omega_{0}}{\Delta\omega_L}
			\left(4 e^{-\chi^2} - 1 - \chi \frac{\Delta\omega_L}{\omega_{0}}\right) \,, 
		\\
	B_1 &= \chi \sqrt{1+B_2^2} \,, 
		\\
	B_0 &= \frac{\chi^2 B_2}{2} - \frac{1}{4} \arctan B_2 + n\pi \,, 
    \end{split}
    \label{B_model}
    \end{align}
    that produce laser pulses which compensate the nonlinear spectrum broadening and significantly reduce the bandwidth of backscattered X-rays.

    \begin{figure}
    \begin{center}
        \includegraphics[width=0.95\columnwidth]{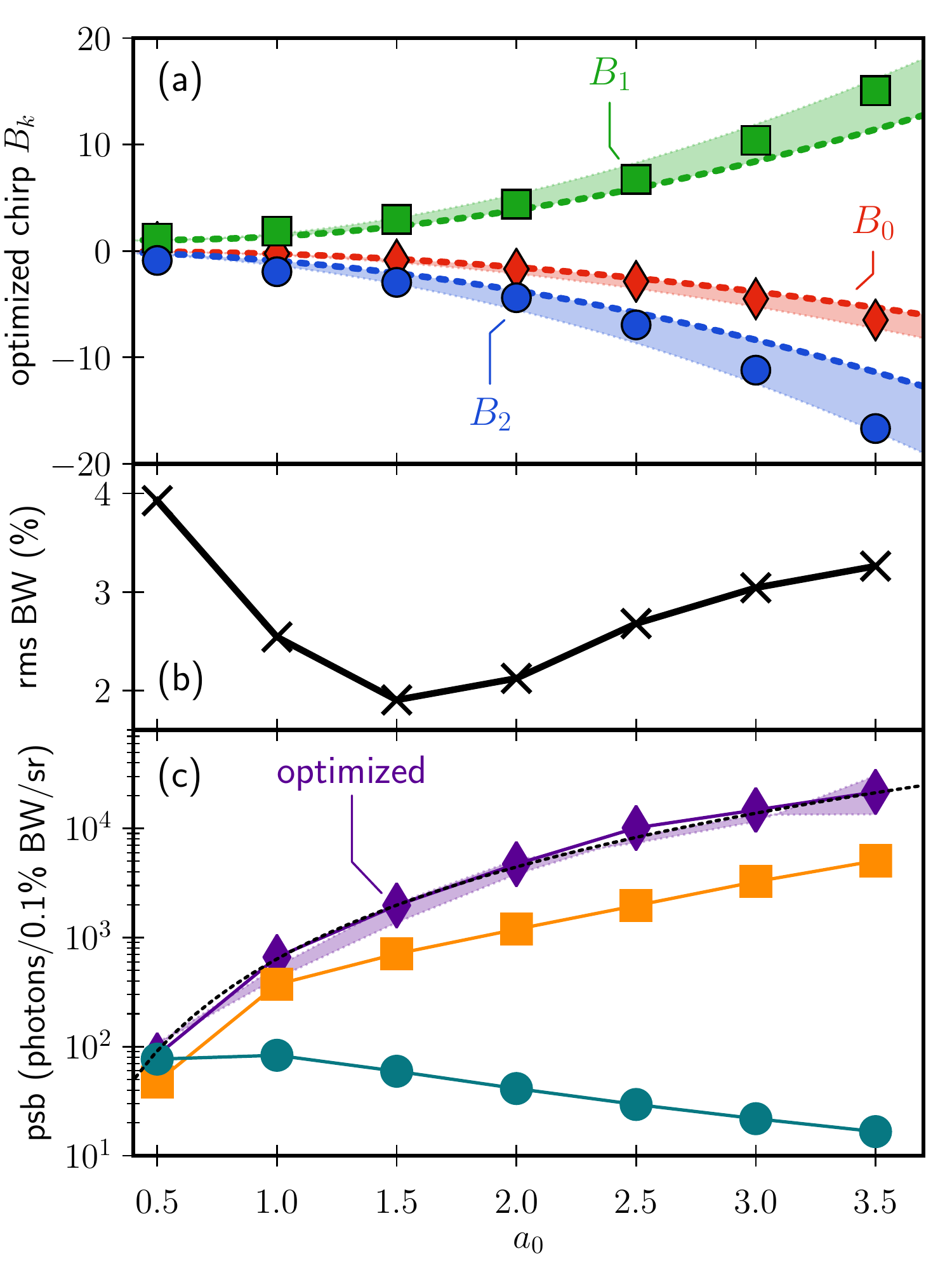}
    \end{center}
	\caption{
	(a) Optimal chirp parameters $B_0$ (red diamonds), $B_1$ (green squares) and $B_2$ (blue circles) that lead to the narrowest on-axis scattered photon spectrum and obtained from simulations, as functions of $a_0$ for $\Delta \omega_L/\omega_0=0.1$. Shaded areas of corresponding colors are the  model predictions, Eqns.~\eqref{B_model}.
    (b) Relative rms bandwidth of the on-axis photon spectrum for the optimally chirped pulse. 
    (c) Peak spectral brightness (psb) of the on-axis scattered photons as a function of $a_0$ for the optimally chirped pulse (purple diamonds) and compared to matched Gaussian (orange squares) and unoptimized cases (cyan circles). The black dotted curve represents a fit $f(a_0)=637\,a_0^{2.8}$ of the optimized case, the shaded are corresponds to peak spectral brightness predicted using model parameters just like in (a).
    }
    \label{sim_fig}
    \end{figure}

    Numerical optimization was carried out in order to find optimal sets of parameters $\vec{ B }$ that yields the narrowest rms~ICS spectrum for various $a_0$ and laser bandwidth $\Delta\omega_L/\omega_{0}=0.1$. The results are shown in Fig.~\ref{sim_fig}, and compared to the analytical model predictions, Eqns.~\eqref{B_model}. The symbols in Fig.~\ref{sim_fig} (a) refer to the numerically optimized parameters $\vec B$, while the shaded areas correspond to the model predictions with varying $\chi=0.95\ldots1$ with the dotted curves at $\chi=1$. Note there are some discrepancies between the numerical optimization and the analytical model for small $a_0$ because we assumed $B_2\gg1$ in order to solve the equations to arrive at \eqref{B_model}.

    Fig.~\ref{sim_fig} (b) shows a relative rms bandwidth of the ICS photons, which is well below 4\% throughout. It is evident there is an optimal range of $a_0$ for the given laser bandwidth around $a_0=1.5$ where the bandwidth of the ICS photons is smallest. This optimal region shifts to higher $a_0$ for larger laser bandwidth.

	For large values of $a_0 > 3.5$, the quality of the photon spectrum somewhat decreases. We attribute this to the fact, that for high values of $a_0$, the amount of chirp stretches the pulses very long and the delay between pulses leads to the beating of two Gaussians such that the envelope of the two-pulse is not smooth anymore. According to our model, for the optimally chirped pulses the effective $a_\mathrm{eff} \approx 1.76 \times \sqrt{ \Delta \omega_L / \omega_0} $ for $|B_2|\gg1$, independent of the initial $a_0$. We have performed studies for different values of $\Delta \omega_L/\omega_0$ and found similar optimization but extending to larger $a_0$ for larger bandwidth.

    Fig.~\ref{sim_fig}(c) shows the on-axis peak spectral brightness of the ICS photons as a function of $a_0$ for the numerically optimized chirped pulses (purple diamonds). The purple shaded area corresponds to the optimized model parameters Eqns.~\eqref{B_model} with the same variation as in Fig.~\ref{sim_fig} (a), showing a very good agreement and robustness. The case of $a_0=2$ is the same as shown in Fig.~\ref{spectra_fig}. Our simulations indicate that the peak  spectral brightness for optimally chirped pulses grows $\propto a_0^{2.8}$ (black dotted curve). For comparison, we also show the completely unchirped pulses (green circles) and matched gaussians (orange squares). The peak spectral brightness for optimally chirped pulse exceeds the matched Gaussian pulse by a factor of $4\ldots5$.

\begin{figure}
    \begin{center}
    \includegraphics[width=0.95\columnwidth]{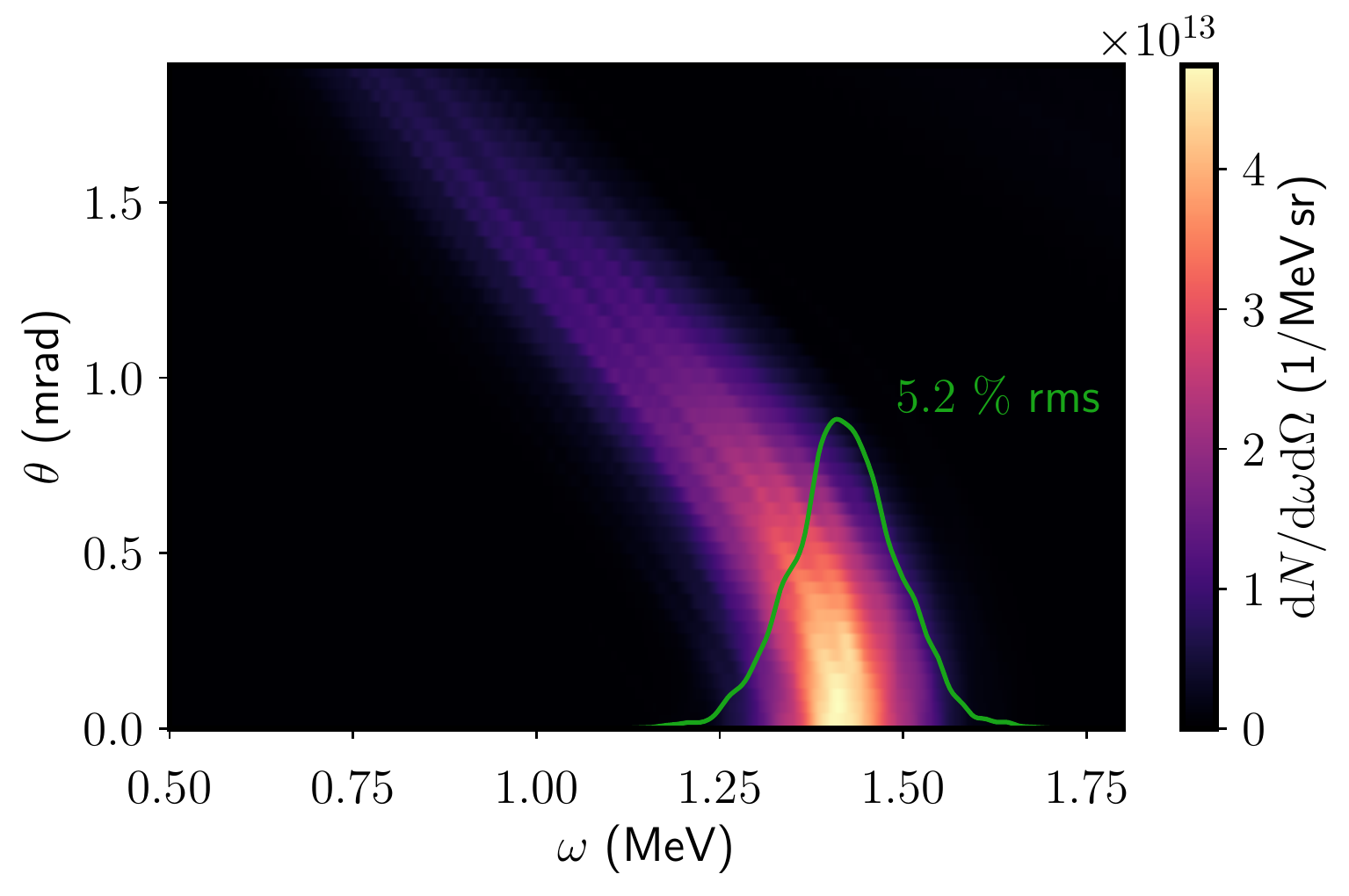}
    \vspace*{-4mm}
    \end{center}
    \caption{\RefA{Simulated spectral-angular Compton photon distribution for a $\unit{50}{\pico\coulomb}$ realistic electron beam interacting with a focused laser pulse. The green curve is an on-axis lineout.}}
    \label{fig:simulation}
    \end{figure}

    \RefA{
    We have presented a simple two-pulse scheme for the compensation of non-linear broadening effects in inverse Compton scattering gamma ray sources based on the spectral synthesis of optimized chirped laser pulses. We developed a model to predict the required spectral phase in dependence on the laser intensity and bandwidth. To verify the robustness of our scheme with regard to 3D effects we numerically simulated electron trajectories and the radiation emission using Li\'enard-Wiechert potentials in the far field \cite{book:Jackson} for a realistic scenario (Figure~\ref{fig:simulation}): A $\unit{270}{\mega\electronvolt}$ LPA electron beam with $\unit{2.2}{\%}$ energy spread and $\unit{0.2}{\milli\metre\,\milli\rad}$ normalized emittance, beam size $\unit{1.8}{\micro\metre}$ and duration $\unit{10}{\femto\second}$ \cite{Lundh:NatPhys2011,Plateau:PRL2012,Weingartner:PRSTAB2012}, is colliding with a Gaussian laser pulse of waist $w_0=\unit{20}{\micro\metre}$ within the paraxial approximation \cite{Harvey:PRSTAB2016,Maroli:JApplPhys2018}, $a_0=2$, $\omega_0=\unit{1.55}{\electronvolt}$, and the temporal chirping taken from the numerical optimization in Fig.~\ref{sim_fig}.
    }

    \RefB{
    Our model predictions can serve as initial conditions for active feedback optimization of inverse Compton sources, e.g.~using machine learning techniques. The optimal control of the temporal laser pulse structure has been successfully demonstrated already, e.g.~for the optimization of laser accelerated electron beams and x-ray production \cite{Streeter:APL2018,Dann:2018}.  The optimal chirping of the scattering pulse could be included into an overall feedback loop. Depending on the desired bandwidth of the
    ICS the other contributions to the total bandwidth, such as the electron energy spread and emittance, could be optimized alongside the temporal spectral shape
    of the scattering laser pulse by choosing proper goal functions for optimization.
    }

	D.~S.~acknowledges fruitful discussions with A.~G.~R.~Thomas and S.~Dann. This work was funded in part by the US ARO grant no.~W911NF-16-1-0044 and by the Helmholtz Association (Young Investigators Group VH-NG-1037).

 \input{references}
\newpage
\begin{widetext}

\setcounter{equation}{0} 
\renewcommand{\theequation}{S.\arabic{equation}}

\part{Supplementary Material}

\section{Analytic Results for the Two-Pulse}

Here we provide explicit formulas for the vector potential of the recombined two-pulse $a_\perp(\varphi)$, its envelope $a(\varphi)$, and local frequency $\omega_L(\varphi)$. First, by calculating the inverse Fourier transformation of the modulated spectrum, 
Eq.~(5) of the main text, 
we immediately find the complex scalar amplitude
\begin{align}
a_\perp & = \frac{a_0}{2 \sqrt[4]{1+B_2^2}} 
e^{-i \omega_0 \varphi }
\left( 
e^{- \frac{c_-}{2} (1+iB_2) + i B_0 + \frac{i}{2} \arctan B_2 } 
+ 
e^{- \frac{c_+}{2} (1-iB_2) - iB_0  - \frac{i}{2} \arctan B_2} 
\right)
\end{align}
with
\begin{align}
c_\pm &= \frac{(B_1 \pm \varphi \Delta \omega_L)^2}{1+B_2^2} \,,
\end{align}
and $\varphi = t-z $.
The real vector potential of a circularly polarized laser pulse is related to $a_\perp$ by 
$\vec a_\perp = \Re [ a_\perp \vec \epsilon ]$ with $\vec \epsilon = \vec e_x + i \vec e_y$.

The squared envelope of the laser pulse is given by
\begin{align} \label{eq:a2}
a^2(\varphi) &= \vec a_\perp^2 = |  a_\perp |^2 
=  \frac{a_0^2}{2\sqrt{1+B_2^2}}  e^{-\xi}
\left[ \cosh 2d_0\varphi
+ \cos \zeta
\right]
\end{align}
where we introduced the following abbreviations
\begin{align}
d_0 & = \frac{B_1 \Delta \omega}{1+B_2^2}\,, \qquad
\xi = \frac{B_1^2 + \varphi^2 \Delta\omega^2_L}{1+B_2^2} \,, \qquad
\zeta = 2 B_0 - B_2 \xi + \arctan B_2 \,.
\end{align}

An important quantity is the infinite integral over the squared vector potential envelope,
\begin{align} \label{total_energy_eq}
\intop_{-\infty}^\infty \! \ud \varphi \: a^2(\varphi) =
\frac{a_0^2\sqrt{\pi}}{2}\frac{1}{\Delta\omega_L}
\left[
1
+
\frac{e^{-\frac{B_1^2}{1+B_2^2}}}{(1+B_2^2)^\frac{1}{4}}
\,
\cos \left( 
{2B_0 - \frac{B_1^2B_2}{1+B_2^2} + \frac{1}{2} \arctan B_2 } \right)
\right] \,.
\end{align}
When the cosine term is maximised, i.e.~its argument a multiple of $2\pi$, then 
the interference between the two sub-pulses is mostly constructive.

Because $a_\perp = a \, e^{- i \Phi(\varphi)}$ we can write
$\Phi = i \log \frac{a_\perp}{a}$, and
\begin{align} \label{eq:omega-analytic}
\omega_L(\varphi) &= \frac{\ud \Phi }{\ud\varphi}   
= -  \frac{{\rm Im}\,[ a_\perp'a_\perp^* ] }{a^2}  
= \omega_{0} - d_0 B_2 + d_0 \: \frac{\varphi \frac{\Delta \omega B_2}{B_1} \sinh 2d_0 \varphi - \sin \zeta}{\cosh 2d_0\varphi + \cos \zeta} \,.
\end{align}

\section{Defining the Matched Gaussian Pulse}

We match the a Gaussian with constant frequency $\omega_0$ to the two-pulse, having the same effective peak amplitude and total energy in the pulse,
\begin{align}
a_\mathrm{matched}(\varphi) = a_\mathrm{eff} \, e^{-i\omega_0\varphi} \, e^{-\varphi^2/2\Delta \varphi_\mathrm{eff}^2}\,.
\end{align}
First, the amplitude is matched by evaluating Eq.~\eqref{eq:a2} at $\varphi=0$ with the approximation $\zeta \to 0$, 
yielding
\begin{align}
a_\mathrm{eff}^2   &= a_0^2 \: \frac{e^{- \frac{B_1^2}{1+B_2^2}}}{\sqrt{1+B^2_2}} \,.
\end{align}

Second, the pulse duration is matched by the requirement that both the chirped two-pulse and the matched Gaussian
have the same energy, 
\begin{align}
w = \int \ud \varphi  \left| \frac{ \ud a_\perp }{\ud \varphi } \right|^2 \,,
\end{align}
i.e. 
\begin{align}
\Delta\varphi_\mathrm{eff} = \frac{w}{\sqrt{\pi} \omega_0^2 a_\mathrm{eff}^2} \,.
\end{align}

\section{Derivation of the Formula for the on-axis Spectrum}

Under the assumption that we can use classical electrodynamics to calculate the radiation spectrum,
the energy and angular differential photon distribution is given by \cite{book:Jackson}
\begin{align} \label{N1}
\frac{\ud N}{\ud \omega' \ud \Omega} = \frac{\omega'}{4\pi^2} | \vec n' \times (\vec n' \times \vec j)|^2 \,,
\end{align}
with the Fourier transformed electron current
\begin{align}
\vec j (\omega',\vec n') =  - e \int \! \ud s \, \vec u(s) \, e^{i \omega' [ t(s) - \vec n' \cdot \vec x(s)]} \,,
\end{align} 
and $\vec n'$ the direction under which the radiation is observed.
Here, $s$ denotes the electron's proper time, parametrizing the electron orbits $t(s),\vec x(s)$ and four-velocity components $\gamma(s) = \ud t/\ud s = \sqrt{1+\vec u^2}$ and $\vec u = \ud \vec x/\ud s$, which is a solution of the Lorentz force equation
\begin{align}
\frac{\ud \vec u}{\ud s} = \frac{e}{m} ( \gamma \vec E + \vec u\times \vec B ) \,.
\end{align}
For on-axis radiation, $\vec n' = - \vec e_z$, the double vector product can be simplified to
\begin{align}
| \vec n' \times (\vec n' \times \vec j)|^2 = | \vec j_\perp |^2 \,,
\end{align} 
and with $\vec u_\perp = -\vec a_\perp$ and changing integration variables from proper time to laser phase
$\varphi = t-z$ via $ \ud \varphi / \ud s \approx  2\gamma $ (for initial value of $\gamma\gg1$), 
\eqref{N1} turns into
\begin{align}
\left.\frac{\ud N}{\ud \omega' \ud \Omega}\right|_\mathrm{on-axis} 
= \frac{e^2 \omega'}{4\pi^2 (2 \gamma)^2} \: \left| \int \! \ud \varphi \, \vec a_\perp \: e^{i\omega' [ t(\varphi) + z(\varphi) ] } \right|^2 \,.
\end{align}
By noting that $t(\varphi) + z(\varphi) = (2\gamma)^{-1} \int \! \ud \varphi [ \gamma(\varphi) + u_z(\varphi) ] $,
with $\gamma(\varphi) + u_z(\varphi) = ( 1 + \vec a_\perp^2)/(2\gamma)$, the definition of the normalized frequency of the emitted photon, $y = \omega' / (4\gamma^2 \omega_{0})$
we eventually arrive at Eq.~(6) of the main text.
In Fig.~\ref{fig:onaxis} we show the on-axis spectra for optimized laser pulses of various intensities and bandwidth $\Delta\omega_L/\omega_0=0.1$.

\begin{figure}
	\begin{center}
		\includegraphics[width=0.6\columnwidth]{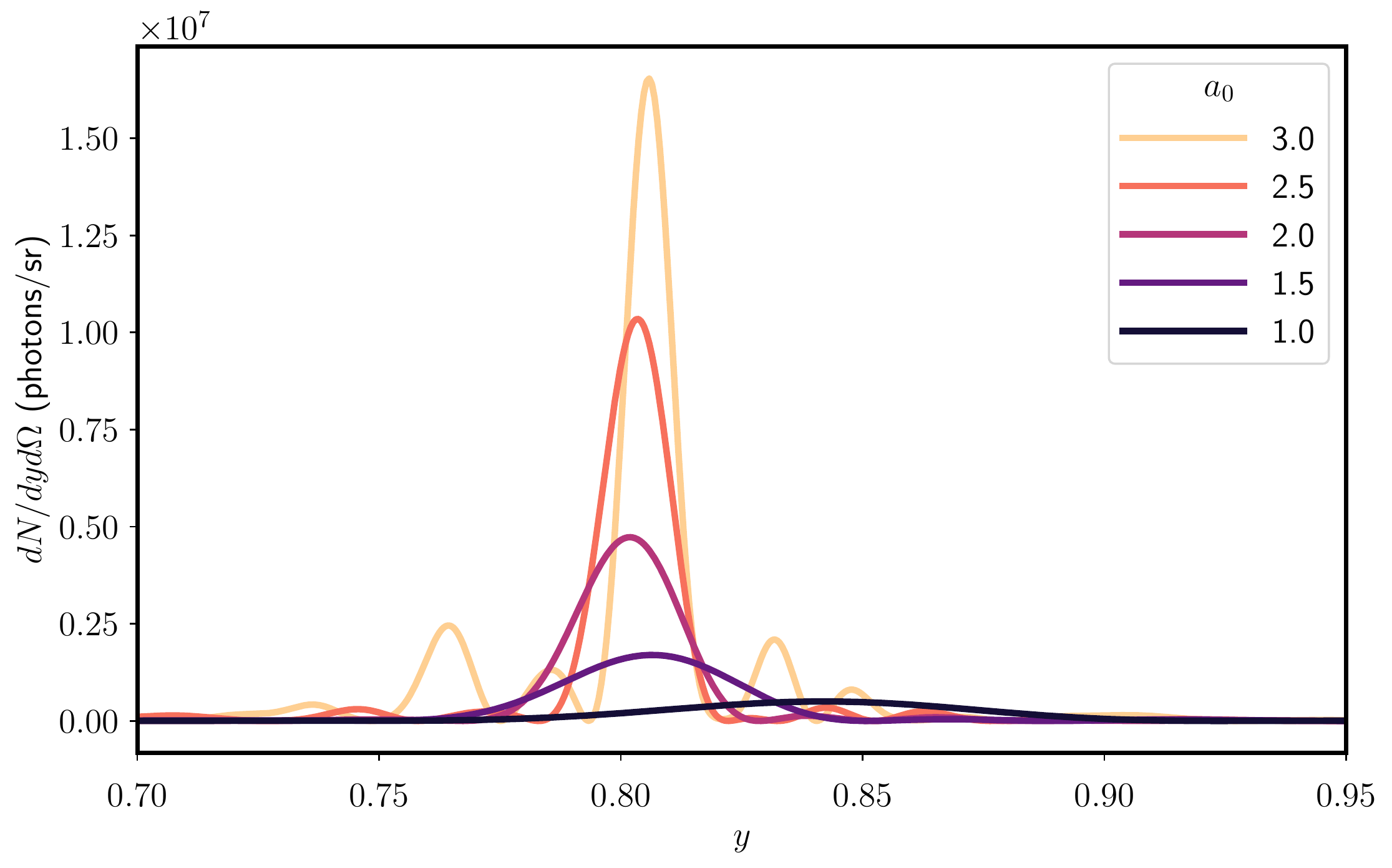}
	\end{center}
	\caption{On-axis photon spectra from optimized laser pulses with the chirping parameters $\vec B$ calculated using the model Eqns.~(8) with $\chi=1$.}
	\label{fig:onaxis}
\end{figure}

\end{widetext}

\end{document}

%% file: references.tex
\providecommand{\noopsort}[1]{}